\def\lanlr#1{\href{http://www.arxiv.org/abs/#1}{#1}}

\documentclass[12pt]{article}%
\usepackage{hyperref}
\begin{document}

    \title{NON-LINEAR QUANTUM MECHANICS AND  HIGH ENERGY COSMIC RAYS}
    
       \author{George  Svetlichny\footnote{Departamento de Matm\'atica, Pontif\'{\i}cia Universidade Cat\'olica, Rio de Janeiro, Brazil \newline
svetlich@mat.puc-rio.br \hfill \url{http://mat.puc-rio.br/\~svetlich}}}

\maketitle
\begin{abstract}
    We point out that the presence of energetic cosmic rays above the GZK cutoff may be explained by fundamental non-linearities in quantum mechanics at the Planck level. 
\end{abstract}
\section{INTRODUCTION}
The apparent presence of highly energetic cosmic rays is seen by many to be enigmatic. The free path of cosmic particles is limited by their interaction with the background thermal radiation and the free path shortens with energy.  As there are no known near sources of very high energy particles, none should be seen above a certain energy cutoff, know as the Greisen-Zatsepin-Kuzmin (GZK) cutoff \cite{gzk}. Observations however  suggest  their presence, though the question is still not entirely settled \cite{blasi}. The presumed presence has been taken as tangible evidence of otherwise observationally unaccessible new fundamental physics, such as quantum gravity or noncommutative geometry \cite{amqg}, though less speculative explanations have also been brought forth \cite{jaimaz}. 

In these theories the presence of high-energy cosmic rays is possible due to a modified momentum-energy dispersion relation with terms beyond the usual quadratic one (here \(p\) is the modulus of the momentum).
\begin{equation}\label{mdr} 
E^2=p^2+m^2+\alpha_3p^3+\alpha_4p^4+\cdots
\end{equation}

Since most fundamental physical theories that transcend the standard model tend to modify the dispersion relation, the presence of cosmic rays above the GKZ limit does not in general single out any particular theory and must be considered as just a general hint of the necessity of {\em  some\/} modification of our fundamental ideas (the observationally acceptable limits on the coefficients \(\alpha_i\) could however {\em  rule out\/} some particular theories).

In this note we wish to point out  that one such possible modification is fundamental 
non-linearity in quantum mechanics. The quadratic dispersion relation is a direct consequence of second order linear theories, and non-linearities would naturally produce higher order terms. Under the proposed conception, classical space-time and low-energy linear quantum mechanics are emergent features of some underlying pre-geometric physical reality. It is now believed that high-energy cosmic rays are possible probes of such a reality. Thus quantum non-linearities, if present,  should  result in non-linear effective wave equations for these particles, and consequently  modified dispersion relations. This conception is an alternative to non-commutative space-time, and we shall point out some of the contrasts below.

\section{QUANTUM NON-LINEARITIES}

 There is by now no dearth of proposals for non-linear quantum mechanics \cite{kos,dg,mavsz,dn} (this is far from an exhaustive  list) and many of these could lead to a non-standard  dispersion relation. However, with the exception of \cite{dn}, these are non-relativistic low-energy theories, first order in the time derivative. It is not clear how relevant they are for the high-energy regime of cosmic rays. In any case one has to compare the dispersion relation of these theories to the version of \(E^2=m^2+p^2\) linear in \(E\) which would be, introducing explicitly the velocity of light \(c\) and disregarding the rest energy,  
\begin{equation}\label{nrde}E=\frac{p^2}{2m}+mc^2\sum_{k=2}^\infty (-1)^{k+1} \frac{1\cdot3\cdots(2k-3)}{2^kk!}\left(\frac{p}{mc}\right)^{2k}.
\end{equation}
One has also to take into account that non-relativistic equations generally do not generate terms beyond the first in (\ref{nrde}) making comparison problematic. If such an equation is the non-relativistic limit of a relativistic equation, then keeping terms of higher order in \(v/c\) should restore the other terms in (\ref{nrde}). Consider now, as an example, 
the Doebner-Goldin \cite{dg} equation 
\begin{equation}\label{dg}
i\hbar\partial_t\psi=-(\hbar^2/2m)\Delta \psi+iD\hbar\left[\frac12
\frac{\Delta \bar\psi\psi}{\bar\psi\psi}\right]\psi+R(\psi)\psi,
\end{equation}
where \(R(\psi)\) is any {\em  real\/} function of homogeneous degree zero (\(R(\lambda\psi)=R(\psi)\)). In particular for \(R(\psi)=f\left(\frac{i\hbar\nabla\psi}\psi\right)\) the dispersion relation would be (assuming \(c=1\) again)
\[E=\frac{p^2}{2m}+f(p)\]
and since \(f\) can be arbitrary (\ref{nrde}) can be modified. This suggests that if the Doebner-Goldin equations can be obtained as  non-relativistic limits of  relativistic equations, the latter generically 
would have a modified dispersion relation (\ref{mdr}). 
The particular choices of \(f\) considered in detail by the authors do not create terms in powers of \(p\) beyond the second and so are not candidates for theories of the type considered here. The Doebner-Goldin equation is particularly interesting as it is related to current algebra which also describes many usual linear quantum systems. Reference \cite{dn} deals with relativistic equations. However these are  mostly equations equivalent to linear ones by  so-called non-linear gauge transformations so these of course have normal dispersion relations. The few examples in \cite{dn} of non-linear equation  not equivalent to linear ones are close generalizations of those that are, and also have standard dispersion relations. On the whole it seems that we lack concrete non-linear theories tailored to  the appropriate high-energy regime of cosmic rays.

Experimental upper bounds on the suppression factor for non-linear effects are of the order of \(10^{-20}\) \cite{nlexp1}. Such small factors are already familiar in fundamental physics. One such is the weakness of gravity in relation to the electro-weak and strong forces, another is the smallness of the cosmological constant. To appreciate the second point consider the Plank time \(\tau=\lambda/c=(G\hbar/c^3)^{1/2}/c=5.39\times 10^{-44}s\) and the present estimate of the cosmological constant \(\Lambda=2.036\times 10^{-35}s^{-2}\). The ratio of 
two times \(\tau/\sqrt \Lambda\) is approximately \(10^{-60}\). We are not suggesting that the suppression  of presumed quantum non-linearities have the same origin as these other effects, only that such small factors are indeed present in the fundamental physical features of our universe. 

As already mentioned, it is now widely accepted that the smooth manifold structure of space-time is an emergent feature of a rather different pre-geometric reality. Whether this be described by quantum gravity, M-theory, or some other more speculative frameworks (see \cite{oodle} for a (already dated) compendium of possibilities) does not matter much for the qualitative consequence of a modified dispersion relation. In such a pre-geometric scenario, there is no compelling reason to think that quantum mechanics be the usual linear one. If we are willing to give up the conventional manifold model of space-time, why insist on the conventional linear model of (quantum) mechanics? Linear quantum mechanics may be just one more emergent feature that a future theory must explain. There are indications that the linearity of quantum mechanics is tied to relativistic causality \cite{cover,svetnl}, so if the latter is an emergent feature, the former could also be.

Many objections have been raised to non-linear quantum mechanics, mainly arguing that such theories violate relativistic causality \cite{svetnl,gisin}. Various proposals have been brought forth to circumvent this \cite{nlrqm}. If however we admit non-linearities at the same level as the emergence of a causal space-time from some pre-geometric reality, then such objections have no force as the causal structure itself is then also not well defined. Non-linearities would become apparent as we probe the pre-geometric regime. As many have already suggested, one possible such probe could be high-energy particles traversing the universe. The emergent space-time may not even be Lorentz invariant, the cosmological frame of isotropic background radiation being privileged. Another possibility is deformed Lorentz symmetry with new rotation and boost transformations without a privileged frame \cite{ha}. In all such  situations, objections based on strict relativistic causality lose their force. Non-linear quantum mechanics in these setting does not need any of the specific mechanisms so far introduce to bring it into conformity with relativity. Similar views have also been expressed by T.~P.~Singh \cite{singh}. It is somewhat suggestive that Mavromatos and Szabo \cite{mavsz} have derived effective non-linear equations for \({\rm D0}\) branes in matrix \(M\)-theory; thus a fundamental linear theory can leads to effective non-linearities. Though our suggestion is the reverse of this, that fundamental non-linearity leads to effective low-energy linear quantum mechanics, it is satisfying to see linearity and non-linearity so related at the Planck level.
 
Now, it is not a priori clear what one should mean by non-linear quantum mechanics. Up to now this has meant using non-linear operators for evolution or measurements in place of the linear ones, and keeping at least some of the usual feature of linear quantum mechanics, such as wave-functions, Hilbert space, use of complex amplitudes, and so on. No one apparently has addressed the problem of quantization in non-linear terms, yet it is this that is needed to address the pre-geometric problem. One can say however that once a manifold space-time and low-energy linear quantum mechanics have emerged, there would be small modifications, stemming from such a different quantization process, to the usual quantum mechanical formulae used to calculate physical quantities. Reciprocally one can say that if one is forced to use such modified calculations one has, {\em  prima facie\/},  passed over to a non-linear theory.

From this point of view the burden of showing plausibility for the non-linear hypothesis is different from that for the other considered alternatives. The modified dispersion relation (\ref{mdr}) is needed to calculate  cross sections for interaction of a high-energy particle with the background radiation. This is normally meant as a linear quantum calculation using a modified plane-wave. One can however read it in a different manner: {\em as an  already non-linear quantum mechanical calculation\/}. In a classical commutative geometry, the dispersion relation (\ref{mdr})  cannot be extracted from a plane wave  \(e^{i(Et-px)}\) by a linear second order differential operator (in non-commutative geometries it's possible \cite{la}), but it can be by a non-linear one. 
Under such a view, one need not show that non-linear quantum mechanics {\em  leads\/} to modified dispersion relations; a modified dispersion relations {\em  is\/} non-linear quantum mechanics. 
Of course this remark relies on speculations about the details of the emergence process leading to classical space-time, second order theories, and low-energy linear quantum mechanics, but at present date no more can be said.

Non-linearities in classical physics lead to many emergent phenomena, solitons and shock waves being among the most striking. It may be that pre-geometric quantum non-linearity is just the ingredient needed to simply and naturally let emerge the world as we know it.

\subsection*{Acknowledgements}
This research was partially supported by the Conselho Nacional de Desenvolvimento Scient\'{\i}fico e Tecnol\'ogico (CNPq).

\end{document}